\documentclass{osa-article}
\journal{oe}
\articletype{Research Article}
\usepackage{multirow,tabularx}
\usepackage{upgreek}
\usepackage{xcolor}
\usepackage{cite}

\newcommand{\doi}[2]
{\href{https://doi.org/#1}{#2}}
\newcommand{\change}[1]{#1}

\begin{document}
\title{Fresnel zone plates for reconfigurable atomic waveguides}

\author{A.~M.~Pike,\authormark{1,$\dagger$}
A.~Dorne,\authormark{1,$\dagger$}
L.~Pickering,\authormark{1}
M.~Jamieson,\authormark{1}
I.~T.~MacCuish,\authormark{1}
E.~Riis,\authormark{1}
M.~Y.~H.~Johnson,\authormark{1}  V.~A.~Henderson,\authormark{1,2} P.~F.~Griffin,\authormark{1} and A.~S.~
Arnold\authormark{1,*}}

\address{\authormark{1}Dept.\ of Physics, SUPA, University of Strathclyde, Glasgow, G4 0NG, UK\\
\authormark{2}RAL Space, STFC, Rutherford Appleton Laboratory, Harwell, Didcot, Oxfordshire, OX11 0QX, UK\\
\email{\authormark{*}aidan.arnold@strath.ac.uk}
\email{\authormark{$\dagger$}These authors contributed equally}}

\begin{abstract}
Fresnel zone plates (FZPs), with patterns of $1\,\upmu$m resolution \change{and $10^{8+}$ `pixels'}, allow the formation of clean, diffraction-limited  foci -- but have a static phase profile. Spatial light modulators (SLMs) allow dynamic control of spatial beam intensity and phase -- but are bulky and currently limited to roughly $10\,\upmu$m pixel sizes and $1\,$Mega-pixel formats. 
Here, we present a new `best-of-both' kind of FZP, scalable to 
\change{rings orders of magnitude larger than those possible via direct SLM generation}. It is equivalent to a plano-convex donut lens, whereby light's local intensity and global phase at the FZP map directly onto the \change{focal} plane. The same FZP under different SLM illumination can generate: rings and arcs, double-rings, phase windings and ring lattices (or dynamic combinations thereof). The smooth and adaptable near-field waveguide this enables will be ideal for Sagnac interferometry with ultracold atoms.
\end{abstract}

\section{Introduction}
Atomic quantum technologies   provide an appealing, accurate, and increasingly portable \cite{McGilligan2022} means of precision sensing for various applications in magnetometry \cite{Ingleby2018,Mitchell2020,Castellucci2021}, timing \cite{mcgrew2018atomic,liu_-orbit_2018,Grotti2018,Elvin2019,Takamoto2020,Martinez2023}, inertial navigation \cite{Gustavson1997,Dutta2016,Menoret2018,Bidel2018,Becker2018,Overstreet2018,aveline_observation_2020,Lee2022,Abend2023}, and fundamental physics \cite{Rosi2014,Morel2020}. The necessary vacuum chambers for atom interferometry techniques are typically spacious to enhance sensitivity via increased interrogation time under finite atomic temperatures and gravity, or for extending the interferometer's enclosed area. Waveguided configurations utilising magnetic \cite{Gupta2005,Arnold2006,Pandey2019,Guo2020}, optical \cite{Beattie2013,Eckel2014,Ryu2020}, or hybrid \cite{Naik2005,Ryu2007,Heathcote2008} potentials for atomic storage  offer a route to reducing the scale of future experiments, and therefore increasing the breadth of environments in which such cold atom sensors can be deployed, such as in compact and mobile sensing platforms or for atomtronic applications \cite{Amico2021}.

Importantly, the loading and propagation of matter waves in any practical waveguide requires smooth potentials. Roughness leads to reflection, heating and fragmentation of Bose-Einstein condensates (BECs) which is detrimental to applications in e.g.\ interferometry -- although isolated controllable potential extrema within a guide can also be utilised as beamsplitters \cite{Marchant2016}, junctions \cite{Ryu2020}, or could be used for delta-kick cooling \cite{Arnold2002,Kovachy2015}.  
Magnetic corrugations seen in chip-based waveguides \cite{Kraft2002,Leanhardt2002} can be removed using fields generated far from the ultracold atoms \cite{Gupta2005,Arnold2006,Pritchard2012,Pandey2019,Guo2020}, however, such fields offer reduced flexibility for creating tunable local barriers within the annular guide, and so many structured-ring experiments are entirely optical. Corrugations in the optical potential caused by interference effects can be minimised using beams spatially filtered by propagating long distances \cite{Kovachy2015}, and subsequent focusing to near the diffraction limit for waveguiding. Even in a smooth waveguide, samples split for interferometry have irreproducible mean-field energy differences which must be mitigated by choosing a species with a tunable scattering length \cite{Boshier2022}, operating at low density or using fermionic species \cite{Cai2022,DelPace2022}. 

While spatial-light modulators (SLMs \cite{Clark2016}) \change{with phase-variable pixels} can be directly used for waveguide creation \cite{Bruce2011}, they suffer from relatively low spatial resolution, aliasing, dead space and cross-talk between pixels \cite{Schroff2023}, and unintentional optical vortices \cite{Gaunt2012}. Digital micromirror devices (DMDs) are another option for generating potentials for  BECs\cite{Gauthier2016}, as they update much faster (tens of kHz compared to tens of Hz with liquid crystal SLMs), but \change{because their pixels are binary in intensity} they have lower efficiency and are still severely limited in spatial resolution compared to FZPs. Similarly acousto-optic modulators and deflectors can create an impressive array of time-averaged 2D patterns \cite{Onofrio2000,Schnelle2008,Houston2008}, but have limited resolution of order $10^4$ spots \cite{Henderson2009} due to the device's finite rise time. Moreover, all optical approaches discussed above rely on the use of a separate `light sheet' to first confine atoms in a plane, whereas FZP potentials should already provide sufficient confinement in the direction of beam propagation \cite{Henderson2020}.

We generate optical waveguides formed by micro-fabricated Fresnel zone plates (FZPs) \cite{Henderson2016}, \change{which}  
allow the creation of a smooth,  optical potential in a compact configuration, as we showed in Ref.~\cite{Henderson2020}. Here, we discuss an evolution of our FZP implementation allowing us to use carefully designed annular static FZPs --- which can then be used in conjunction with dynamically updated illumination. This design allows both error correction of the final fields, as well as greatly expanded control over the generated optical potentials by a clear mapping of intentional angular and/or radial variation in illumination from the input FZP plane to the output focal plane.

\section{Methods: Theory for locally-addressable rings}

\begin{figure}[!b]	\centering
\includegraphics[width=\columnwidth]{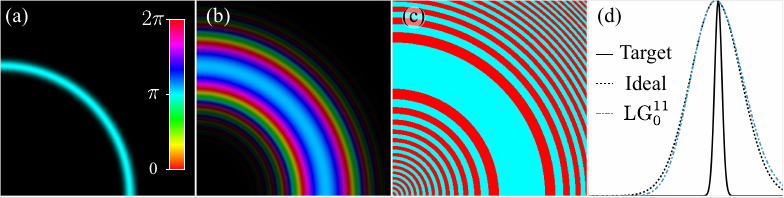}
	\caption[]{\label{fig:mapping} \change{\textbf{Locally-addressable FZP design process --}} illustrated in one ring quadrant with side length $375\,\upmu$m \change{for (a)-(c)}. The simplest target focal electric field is depicted, with radius $r_0=250\,\upmu$m and waist $w=10\,\upmu$m (a). \change{In images (a)-(c) the phase $0-2\pi$ is shown by the hue (inset colorbar in (a)) whilst linear intensity (black to bright) depicts amplitude. The target in (a)} is Fourier-propagated backwards in space  \cite{Henderson2016,Henderson2018,Henderson2020} (b), to show the ideal field illumination at the plane of the FZP. The phase can be digitised to a 2-level (or 4-level) pattern (c) which can then be manufactured. The final image (d) shows the relative radial field amplitude  for the target (\change{black}), the ideal illumination at the zone plate (\change{black-dashed}) and a Laguerre-Gauss LG$_0^{11}$ beam (\change{blue} dash-dotted with waist $100\,\upmu$m) for comparison.}
\end{figure}

In our previous work on micro-fabricated binary FZPs \cite{Henderson2016,Henderson2020}, any point at the pattern's target focal plane was due to constructive interference from all points of a laser beam across the entire surface of the FZP. 
Moreover, circular intensity patterns suitable for ring traps had an experimentally measured RMS error of 3\% in the trapping region \cite{Henderson2020}. 
Here, we design FZPs that mainly provide radial focusing, arising from addressing a limited annular region on the FZP -- effectively creating a donut lens. \change{Note that, apart from this new optical concept, the mathematical methods used to design the FZPs are very similar to those detailed in our publications \cite{Henderson2016,Henderson2018,Henderson2020,MattThesis}, including simple scaling laws for  ring focal intensities and other parameters generalisable to other atomic species (and e.g.\ optical tweezers for biology) \cite{Henderson2020}}.

Fig.~\ref{fig:mapping} illustrates the FZP design process for locally addressable rings. We start with the electric field of our target pattern \change{(Fig.~\ref{fig:mapping} (a))}, which is a ring with Gaussian radial profile of focal waist $w$ ($1/e$-field radius) centred on a radius $r_0$. The target field is propagated backwards in space \change{(Fig.~\ref{fig:mapping} (b))}, by a distance small enough that the approximately `Gaussian' radial profile at the FZP plane has a waist smaller than the ring radius -- but also a waist large enough so that the zone plate has enough radial phase steps for good lensing. In practise we chose a waist at the FZP plane of  $w_\textrm{FZP}\approx 0.28 r_0$, which is centred at a peak intensity radius $r\approx 0.98 r_0$ (due to the cylindrical rather than cartesian symmetry), for all FZPs we manufactured. This ideal spatial illumination intensity, relative to the ring radius, is approximately the same even for different focal waists $w$, however the phase variation differs significantly as $w$ is inversely proportional to the annular beam's angular convergence after the zone plate. 

Because of the FZP generation process the focal length $f$ is proportional to both the ring radius and waist -- our $1.07\,\upmu$m wavelength FZPs have $f \approx r_0  w/(1.25\,\upmu\textrm{m}).$ This scale factor $1.25\,\upmu\textrm{m}$ was chosen to avoid high focal numerical apertures (specifically, our maximum $1/e^2$ focused intensity contour angle is limited to $<0.13\,$rad), however higher numerical apertures in conjunction with vector light could open interesting applications \cite{Svensson2025}. 
Moreover, as part of the design process, we terminate radial phase jumps on the zone plate when the optimal illumination intensity drops to $e^{-6}\sim 0.2\%$ of its maximal value (not shown in Fig.~\ref{fig:mapping}(c)), preventing limitations to fabrication from  resolution requirements.

Sixteen different FZPs for rings were designed on the same anti-reflection coated substrate. The four radii were $r_0\in\{0.25,\,0.50,\,1.00,\,2.00\}\,$mm, and for each radius there were target radial beam waists of $w\in\{2.5,\,2.5,\,5.0,\, 10.0\}\,\upmu$m. One of the 2.5$\,\upmu$m waist FZPs was a two-level (binary) pattern, and the other three were four-level patterns. The four-level patterns can reach diffraction efficiencies around $80\%$, whereas the binary patterns are limited to $40\%$\change{, i.e.\ similar to the efficiencies achievable with SLMs and DMDs, respectively}. During the FZP design process it became clear that both the diffraction efficiency \change{(e.g.~$79\,\%-83\,\%$)} and rms intensity smoothness \change{(e.g.~$1\,\%-2\,\%$)} of the ring patterns are affected by the choice of where digitisation occurs -- i.e.\ which phase offset starts the continuous -- modulo-$2\pi$ -- phase range. All FZPs fabricated had a phase offset optimised for both smoothness and efficiency \change{-- which were often correlated -- and binary patterns had larger intensity errors.} The radial waist compression factor during focusing is $\times7$ in our simplest ring (Fig.~\ref{fig:mapping}), and $\times 222$ for \change{the} FZP with $r_0=2000\,\upmu$m and waist $w=2.5\,\upmu$m.

\section{Results: experimental locally addressable rings}
\begin{figure}[!b]
	\centering
\includegraphics[width=\textwidth]{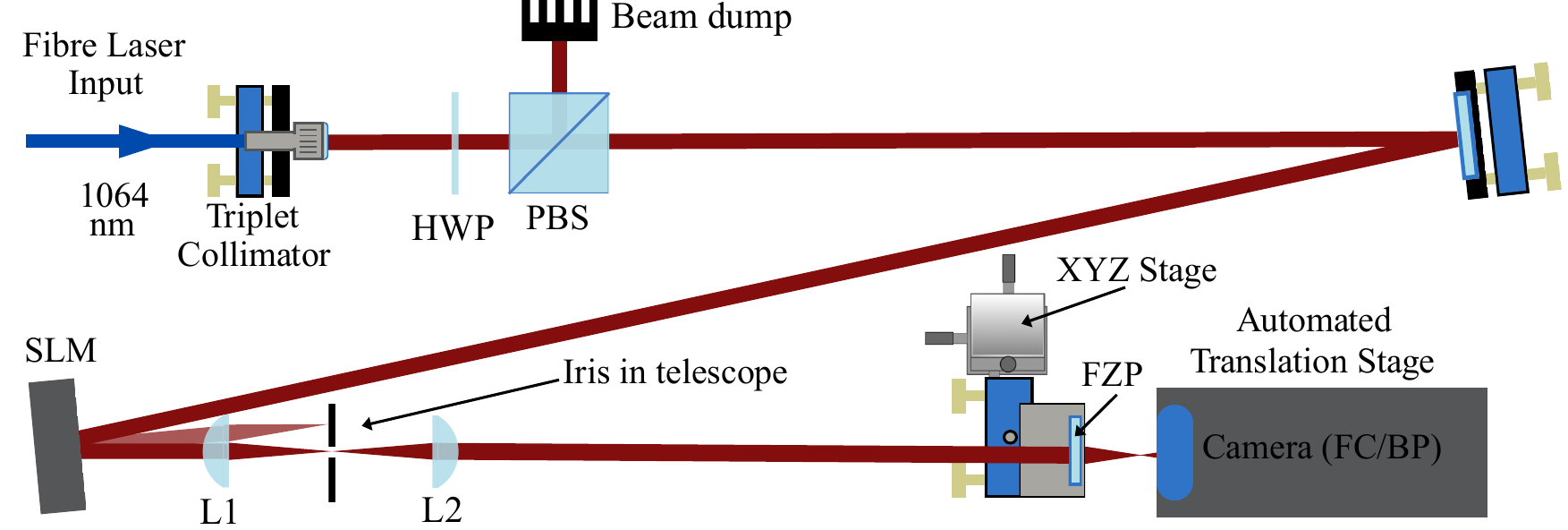}
\caption{\label{figsetup} \change{\textbf{Simplified experimental setup}.}  \change{A collimated Gaussian $1064\,$nm beam  is attenuated by a half-wave plate (HWP) and a polarising beamsplitter (PBS). The SLM shapes the beam, and the first diffracted order is selected by an iris before the FZP. 
A mirror mount and XYZ stage allow FZP alignment. A camera on an automated translation stage is used to study the FZP's focal behaviour. L1 and L2 are $\sim20\,$cm focal length lenses. Cameras used are the FC: FLIR Chameleon USB 2 camera, BP: Cinogy CMOS-1201 beam profiler.}} 
\end{figure}

The experimental setup used to test the rings is depicted in Fig.~\ref{figsetup}. 
An SLM \change{(Hamamatsu X13138-03)} allowed us to shape the light illuminating the FZP, yielding high-quality spatial intensity distributions as shown in Refs.~\cite{Clark2016,Offer2018}. Because it has the longest focal length, our results here utilise the largest FZP ring diameter $2 r_0=4.0\,$mm, with the widest waist $w=10\,\upmu$m. This allows direct imaging of the FZP-generated ring foci with $3.75\,\upmu$m and $5.2\,\upmu$m pizel resolution using an FLIR Chameleon USB 2 camera (FC) and Cinogy CMOS-1201 beam profiler (BP), with sensor dimensions $3.6\,$mm$\times4.8\,$mm and $5.3\,$mm$\times6.7\,$mm, respectively. Measured ring waists in the focal plane -- with the same illumination -- were smaller using the FC instead of the BP, which we attribute to the higher degree of pixelation and/or an image sensor coating leading to pixel cross-talk on the BP. Whilst direct imaging of the other FZPs is not possible, this could be achieved using a suitable lens system to optically relay the FZP focal plane further from the zone plate.

\begin{figure}[!b]\centering\includegraphics[]{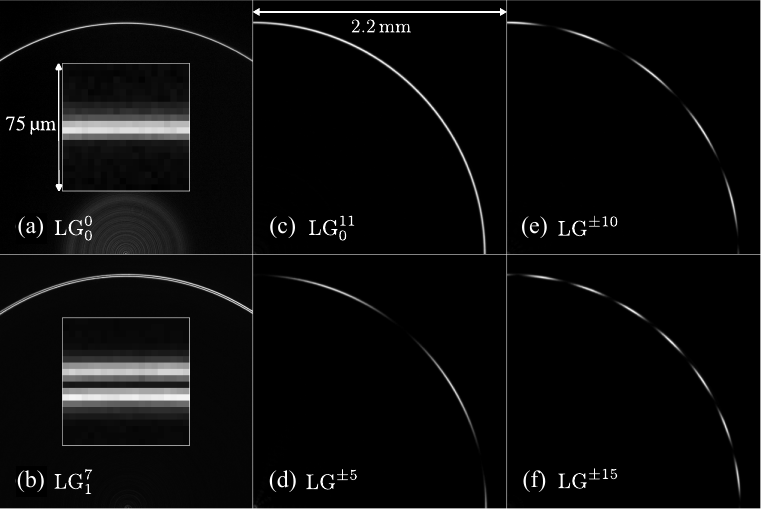}
	\caption[]{\label{fig:lattice}\change{\textbf{The same FZP under different illumination}.} Each experimental image has a linear intensity scale, an image size ($2.2\,$mm)$^2$, and ring radius $r_0\approx2.0\,$mm. The input light is: (a) large waist Gaussian (LG$_{p=0}^{\ell=0}$); 
    (b) LG$_{1}^{7}$; 
    (c) LG$_{0}^{11}$ (i.e.\ similar to Fig.~\ref{fig:mapping}(d)); (d) LG$^{\pm5}$; (e) LG$^{\pm10}$; (f) LG$^{\pm15}$. The insets in (a) and (b) show $(75\,\upmu$m$)^2$ sections from the top of the ring and double-ring, respectively. Images (a)-(b) were taken with the FC for higher spatial resolution, and (c)-(f) using the BP for whole-ring imaging. The corresponding fitted beam waists (a-f),  are $8.0(3)\,\upmu$m, $9(1)\,\upmu$m,  $11.6(5)\,\upmu$m, $11.5(7)\,\upmu$m, $11.6(8)\,\upmu$m, and $11.8(7)\,\upmu$m, respectively. For  (c)-(f), where we use ring-shaped illumination, the beam power fractions in the region $r_0\pm 20\,\upmu$m compared to the overall power in the full image is high 
    ($97\,\%$, $89\,\%$, $97\,\%$ and $97\,\%$, respectively).}
\end{figure}

We have illuminated the \textit{same} FZP with a wide variety of input transverse beam profiles centred on the zone-plate (Fig.~\ref{fig:lattice}), including Laguerre Gauss (LG$_p^{\ell}$) beams \cite{Clark2016,Offer2018}. In particular we have used: (a) a large-waist Gaussian LG$_{0}^{0}$ to overfill the FZP but make a ring; (b) LG$_{1}^{7}$ for a radial double-ring; (c) LG$_{0}^{11}$ for a ring; as well as superpositions of LG beams or `optical ferris wheels' \cite{FrankeArnold2007,Arnold2012} LG$^{\pm\ell}=($LG$_0^\ell+$LG$_0^{-\ell})/\sqrt{2}$ in (d) LG$^{\pm5}$; (e) LG$^{\pm10}$ and (f) LG$^{\pm15}$ to make azimuthal ring lattices. In this paper, light is shaped using single-pass Gaussian illumination of an SLM, however much higher total conversion efficiencies could be achieved using a multi-pass SLM geometry \cite{Fontaine2019,Scholes2020}.

\begin{figure}[!b]	\centering
\includegraphics[]{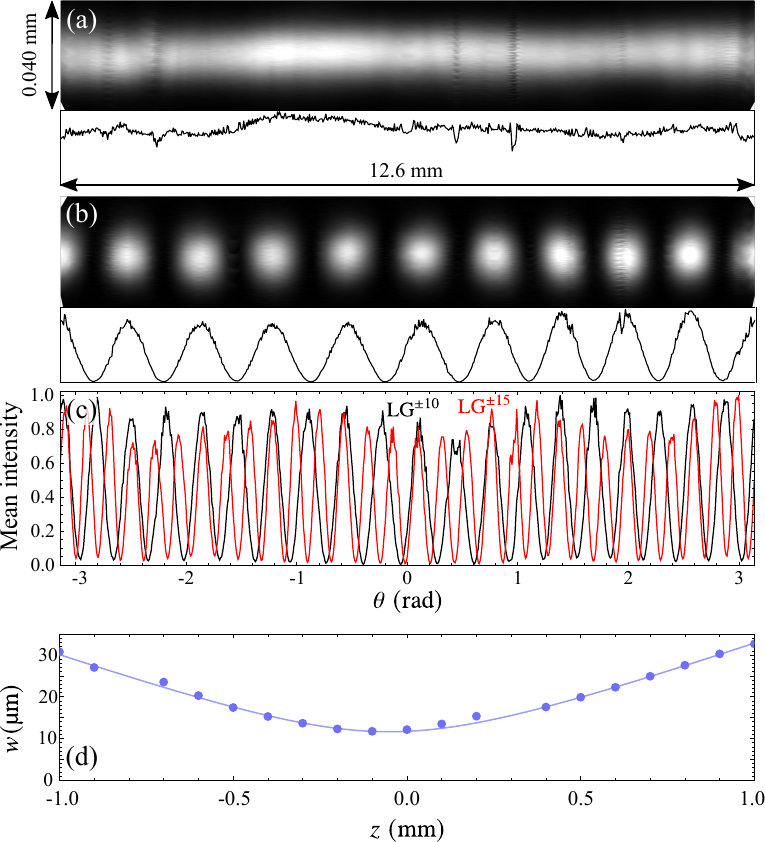}
	\caption{\label{fig:unwrap} \change{\textbf{Cylindrical polar co-ordinate beam projections, with axial propagation.}} In (a) and (b), we show more detail from the FZP ring focal plane images  Fig.~\ref{fig:lattice} (c) (LG$_{0}^{11}$) and (d)  (LG$^{\pm5}$). In (c) angular plots highlight the ring lattices associated with Figs.~\ref{fig:lattice}(e)-(f) (LG$^{\pm10}$ and  LG$^{\pm15}$), in black and red, respectively. Beam propagation can lead to axial confinement in rings and ring lattices, as shown in (d) where the fitted beam waist, averaged around the ring lattice, from Fig.~\ref{fig:lattice} (e) is plotted as a function of camera position.}
\end{figure}

We now fully illustrate  that we can create rings and ring-lattices that can provide both tight azimuthal (Fig.~\ref{fig:unwrap} (a-c)) and axial confinement (Fig.~\ref{fig:unwrap} (d)). This is illustrated by `unwrapping' the uncropped FZP focal plane images from Fig.~\ref{fig:lattice} (c-f) into polar co-ordinates, as well as by mapping the radial intensity distribution as a function of position along the laser beam, using the automated translation stage in Fig.~\ref{figsetup}. 

\change{In Fig.~\ref{fig:unwrap} (a) and (b)
the main images are $40\,\upmu$m high and $12.6\,$mm wide -- i.e.\ we have exaggerated the radial aspect ratio by a factor of 50. Artefacts in the images mainly arise due to the cartesian-to-polar coordinate transform. In the lower halves of (a) and (b), as well as graph (c), we show the corresponding angular variation of the mean intensity in the radial region $r=r_0\pm7\upmu$m. The fit showing behaviour through the focus in Fig.~\ref{fig:unwrap} (d) assumes 1D propagation of a Gaussian beam with waist $11.6\,\upmu$m, with very similar plots arising from Figs.~\ref{fig:lattice} (c-f). Note all data here corresponds to our `worst' design focal waist of $10\,\upmu$m, and a waist of $2.5\,\upmu$m reduces the trap size  $4\times$ and $16\times$  in the radial and axial directions, respectively.}

In this paper we have so far mainly covered various possibilities for trapping atoms within bright regions of red-detuned light. However, trapping ultracold atoms in regions of low intensity using blue-detuned light allows the reduction of photon scattering and thereby heating\cite{Arnold2012,Gaunt2013,Hannover2025}. Such traps can be formed from an optical potential consisting of two rings with opposite phase and different radii, forming an intensity minimum at the midpoint of their radii and providing radial confinement (cf.~Fig.~\ref{fig:lattice} (b)). This minimum will even be a zero crossing of the electric field itself, and thus fully dark. This ensures the waveguide necessarily consists of a zero-intensity central region that cannot vary azimuthally as the atoms move around the ring. 

\begin{figure}[!b]
\centering
\includegraphics[]{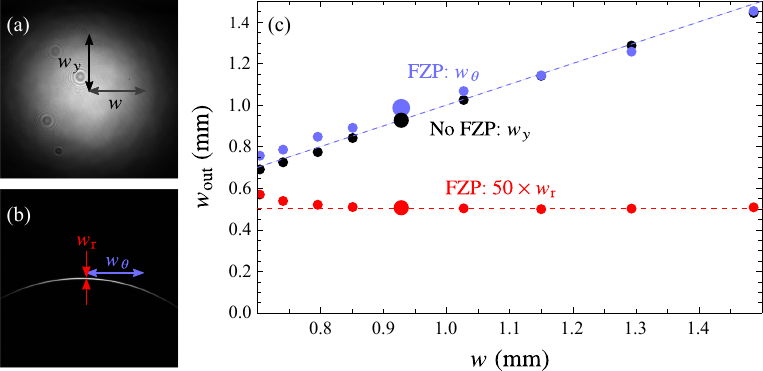}
\caption{\label{fig:local} \change{\textbf{Demonstration of local addressing.}} A Gaussian beam with a waist smaller than the ring radius locally illuminates the FZP (a), and then propagates to the focal plane (b) -- focusing almost entirely in the radial direction. In (c) the unfocused radial waist \change{$w_y$} (black), azimuthal  waist $w_\theta$ (blue) and 50 times the radial waist $w_r$ (red - where $0.5\,$mm$\,=50\times10\,\upmu$m) are shown as a function of the horizontal input beam waist, \change{$w$} ($x$-axis). The larger dots correspond to the data used for (a) and (b).}
\end{figure}

These double-ring potentials \cite{Arnold2012,Turpin2015} can be formed in our hybrid setup using the existing, versatile FZP with appropriate illumination. A double-ring intensity distribution can be projected onto the FZP, which then maps to the focal plane creating a double ring. If, rather than using modes of the electric field (Fig.~\ref{fig:lattice}), we re-image patterns from the SLM plane at the zone plate, we can use an SLM to impart a radial $\pi$ phase slip across the input field, preserving the sign of the phase across the double-ring illumination. We predict this should yield a dark ring localised in 3D, as a counterpart to the bright ring of Fig.~\ref{fig:unwrap} (d). This would therefore eliminate the need for additional `light sheet' beams entirely for both dark and bright ring traps by careful construction of the illumination fields. 

Only the transfer of electric field topology has been considered so far, i.e.\ azimuthal and radial electric field nodes (in the form of ring lattices and double rings) have mapped from the zone plate to the focal plane. As a final illustration of the versatility of these ring FZPs, we now demonstrate local addressability by illuminating the FZP with a Gaussian beam of variable spot size (Fig.~\ref{fig:local}). For a wide variety of illuminating waists there is very little focussing in the azimuthal direction, however, there is strong radial focusing to the $10\,\upmu$m target waist. Illuminating the FZP with a ring of light with azimuthally varying intensity will therefore map into a corresponding  azimuthally variation in intensity at the focal place, which can therefore be used as a first step in improving the actual focal illumination with the target.

\section{Conclusions}

In this paper we have demonstrated a new kind of FZP optic exploiting the ability to locally map  input field to the focal plane of the FZP. This 
configuration allows dynamic updating of the FZP illumination -- enhancing the versatility of static FZP optics without compromising on performance.
Both simulations and optical characterisation demonstrate  this local mapping, enabling both bright and dark rings in 3D, as well as ring lattices, all from a single FZP. The design wavelength of the FZP just has to be tailored to either red- or blue-detuning if bright or dark potentials are desired. Moreover, the geometry enables simple output error correction. 
This hybrid system provides a promising route to  versatile and high performance optical waveguides in compact experimental configurations \change{with potential for Dammann enhancement \cite{Anton2024}}.

Whilst optical FZP characterisation is useful, ultracold atoms are the ultimate means to look for any intensity corrugations, as they are highly sensitive to the associated ring guide potential energy landscape. We are in the final stages of implementing these atomic tests experimentally, with ultracold $^{87}$Rb atoms already loaded into a ring dipole trap from the $r_0=250\,\upmu$m FZPs \cite{NextOne}.

\section*{Funding}
Defence Science and Technology Laboratory. Engineering and Physics Sciences Research Council (EPSRC) EP/X012689/1.

\section*{Acknowledgement}
In addition to the support provided by the funding organizations, we are grateful for valuable SLM discussions with Rachel Offer and Andrew Daffurn.

\section*{Disclosures}
The authors declare no conflicts of interest.

\section*{Data Availability}
Data underlying the results presented in this paper are available in Ref.~\cite{Dataset}. For the purpose of open access, the authors have applied a Creative Commons Attribution (CC BY) licence to any Author Accepted Manuscript (AAM) version arising from this submission.

\bibliography{Bib.bib}

\end{document}